\documentstyle[12pt,twoside,fleqn,espcrc1,amsfonts]{article}

\newcommand{\sd}{|\downarrow\rangle}
\newcommand{\su}{|\uparrow\rangle}
\newcommand{\AmS}{{\protect\the\textfont2
  A\kern-.1667em\lower.5ex\hbox{M}\kern-.125emS}}

\title{Quantum Computation, Spectroscopy of Trapped Ions,
\newline and Schr\"odinger's Cat
\thanks{Contribution of NIST; not subject to U.S. copyright}}

\author{D.J. Wineland, C. Monroe, W.M. Itano, D. Kielpinski, B.E. King, C.J. Myatt,
Q.A. Turchette, and C.S. Wood \linebreak \linebreak
National Institute of Standards and Technology (NIST), Boulder, CO,
80303}

\begin{document}
\maketitle

\begin{abstract}
We summarize efforts at NIST to implement quantum computation using trapped ions,
based on a scheme proposed by J.I. Cirac and P. Zoller (Innsbruck
University).  The use of
quantum logic to create entangled states, which can maximize the
quantum-limited signal-to-noise ratio in spectroscopy, is discussed.
\end{abstract}

\section{INTRODUCTION}

The invention by Peter Shor \cite{shor94} of a quantum algorithm for factorizing large
numbers has stimulated a host of theoretical and experimental investigations in the
field of quantum information \cite{q_comp_gen}.  In the area of quantum computation,
various schemes have been proposed to realize experimentally a model
quantum computer \cite{q_comp_gen}.  In the ion storage group at NIST,
we are trying to realize such a device based on the proposal by Cirac
and Zoller \cite{cirac+95}.

In the Cirac-Zoller scheme, qubits are formed from 
two internal energy states, labeled $\sd$ and $\su$, of trapped atomic ions.
If the ions are laser cooled in the same trap, they
form a crystalline array whose vibrations can be described in terms of
normal modes.  The ground and first excited states of a selected mode
can also form a qubit.  This qubit
can serve as a data bus, since the normal modes are a {\em shared} property of
the ions.  An individual ion in the array can be 
coherently manipulated and coupled to the selected normal mode
by using focused laser beams \cite{cirac+95}.  A universal logic
operation, such as a controlled-not (CN) logic gate between
ion qubit $i$ and ion qubit $j$, is accomplished by (1) mapping
the internal state of qubit $i$ onto the selected motional qubit,  (2) performing a CN
between the motional qubit and qubit $j$, and (3) mapping the motional qubit
state back onto qubit $i$.  Each of these steps has been accomplished in the
NIST experiments with a single ion \cite{monroe+gate,wineland+bible}.
We are currently devoting efforts to: (1) scaling quantum logic
operations to two or more ions (Sec.~\ref{sec-experiments}), (2) applying
quantum logic to study fundamental measurement problems on EPR and
GHZ-like states, and (3) applying quantum logic to fundamentally improve the
signal-to-noise ratio (SNR) in spectroscopy and atomic clocks.  In this paper
we briefly discuss this last application.  We are aware of
similar efforts to implement trapped-ion quantum logic at
IBM, Almaden; Innsbruck University; Los Alamos National Laboratory;
Max Planck Institute, Garching; and Oxford University.

\section{ENTANGLED STATES FOR SPECTROSCOPY}

A collection of atoms (neutral or charged) whose internal states are entangled
in a specific way can improve the quantum-limited
SNR in spectroscopy.  This application of quantum logic to form entanglement
is useful with a relatively small
number of atoms and logic operations.  For example, for high-accuracy,
ion-based frequency standards \cite{berkeland+98}, a relatively small number
of trapped ions ($L \leq 100 $) appears optimum due to
various experimental constraints; with $L = 10 - 100$,
a significant improvement in performance in atomic clocks could be
expected.  In contrast, factoring a number which cannot easily be
factored on a classical computer would require considerably more
ions and operations.

In spectroscopy experiments on $L$ atoms, in which the observable is atomic population,
we can view the problem in the
following way using the spin-1/2 analog for two-level atoms.  The total
angular momentum of the system is given by
${\bf J}  =  \sum_{i=1}^{L} {\bf S}_i $, where ${\bf S}_i$
is the spin of the $i$th atom ($S_i = 1/2$).  The task is to measure $\omega_{0}$,
the frequency of transitions between the $\sd$ and $\su$ states,
relative to the frequency $\omega_R$ of a reference oscillator.
We first prepare an initial state for the spins.   Typically, spectroscopy
is performed by applying (classical) fields of frequency $\omega_{R}$
for a time $T_{R}$ according to the method of
separated fields by Ramsey \cite{ramsey_book}.
We assume the same field amplitude is applied to all atoms (the phases might be
different) and that the maximum value of
$T_R$ is fixed by experimental constraints
(Sec.~\ref{sec-applicability}).  After applying these fields,
we measure the final state populations; for example,
the number of atoms $L_{\downarrow} $ in the $\sd$ state.
In trapped-ion experiments, this has been accomplished through
laser fluorescence detection with nearly 100\% efficiency, which we
assume here
(see the discussion and
references in Ref. \cite{wineland+bible}).
In the spin-1/2 analog, measuring  $L_{\downarrow} $
is equivalent to measuring the operator $J_{z}$,
since $L_{\downarrow} = J$$\Bbb{I}$$ - J_{z}$
where $\Bbb{I}$ is the identity operator.  The SNR (for repeated measurements)
is fundamentally limited by the quantum fluctuations in the number of
atoms which are observed to be in the $\sd$ state.   These fluctuations
can be called quantum projection noise \cite{itano+93}.  Spectroscopy is
typically performed on $L$ initially unentangled atoms (for example,
$\Psi(t=0) = \prod_{i=1}^{L} \sd_{i}$) which remain unentangled
after the application of the
Ramsey fields.  For this case, the imprecision
in a determination of the frequency of the transition is limited by
projection noise to
the ``shot noise" limit $(\Delta \omega)_{meas} = 1/\sqrt{LT_{R} \tau }$
where $\tau \gg T_{R}$ is the total averaging time \cite{itano+93}.
If the atoms can be prepared initially in particular entangled states,
it is possible to achieve $(\Delta \omega)_{meas} < 1/\sqrt{LT_{R} \tau }$.

In optics, squeezed states have been shown to improve the SNR in
interferometers
beyond the shot noise limit \cite{caves81,xiao+87}.
In 1986, Yurke \cite{yurke86} showed how particular entangled states, if
they could be created,
could be used as inputs to Mach-Zehnder interferometers to approach the Heisenberg
limit of SNR.
In 1991, Kitegawa and Ueda \cite{kitegawa+91} showed how the Coulomb
interaction between electrons in the two arms of an electron
interferometer might be used to improve the SNR beyond the shot-noise
limit.  Because of the formal identity of Mach-Zehnder interferometers
and Ramsey spectroscopy \cite{wineland+92}, similar ideas might be
applied to the spectroscopy problem.  Reference \cite{wineland+92}
showed how a Jaynes-Cummings-type coupling between trapped-ion internal states
and a normal mode could be used to improve the SNR in
spectroscopy beyond the shot-noise limit.  The scheme in Ref. \cite{wineland+92}
has the advantage
that the appropriate states can be generated by acting on all the ions
at once (thus not requiring focused laser beams), but has the
disadvantage that these states are entangled with the motion, thereby
requiring small motional decoherence.  Reference \cite{bollinger+96}
investigated the use of the generalized GHZ
state, sometimes called the maximally entangled state,
in spectroscopy.  This state has the form

\begin{equation}
\psi_{max} = {1\over\sqrt{2}}\Biggr(\sd_1\sd_2\cdots\sd_L
+ e^{i\phi(t)}\su_1\su_2\cdots\su_L\Biggl),
\label{psi-max} 
\end{equation}
where $\phi (t) = \phi_0 - L \omega_0 t $.  After application of the Ramsey radiation,
we measure the operator $\tilde{O} \equiv \prod_{i=1}^L S_{zi}$.
The resulting signal gives
the exact Heisenberg limit of SNR
($(\Delta \omega)_{meas} = 1/L \sqrt{T_{R} \tau }$
where $\tau \gg T_{R}$)
in spectroscopy (and
interferometry).

The state $\psi_{max}$
can be generated in a straightforward way by the application of $L$ CN
gates \cite{cirac+95}.  An alternative method was suggested in Ref.
\cite{bollinger+96} and in Refs. \cite{wineland+bible} and
\cite{steinbach+98} methods to generate $\psi_{max}$ with a fixed number of
steps (independent of $L$) are discussed.  For all of these methods, the
the motion is entangled with internal states during the
creation of $\psi_{max}$, but is not entangled afterwards.  Therefore,
once $\psi_{max}$ is created, the motion can lose coherence without affecting the
entanglement of the internal states.

\subsection{Schr\"odinger's Cat}

As $L$ becomes large and more macroscopic, states like $\psi_{max}$ become more like
Schr\"odinger's cat in that they represent coherent superpositions between widely
separate regions of a large Hilbert space; for example, \mbox{$\su_1\su_2\cdots\su_L$} 
$\Longleftrightarrow$ ``live cat;" 
\mbox{$\sd_1\sd_2\cdots\sd_L$}
$\Longleftrightarrow$ ``dead cat".  As has been emphasized in many
discussions, as $L$ becomes large the coherence between the two components
of the cat becomes harder and harder to preserve \cite{haroche98}.  
This is apparent in Eq. (\ref{psi-max}) because if, for example,
$\omega_0$ fluctuates randomly, the two components of $\psi_{max}$ will
decohere relative to each other $L$ times
faster than for one ion ($\psi_{max}$ for $L=1$).
Trapped ions are interesting because it may be possible to make $L$ very large
without significant decoherence.  This is the same property that makes
trapped ions interesting as possible frequency standards.  For example,
in Refs. \cite{bollinger+91} and \cite{fisk+95}, coherence times
for individual ions ($L = 1$) exceeding
10 minutes were obtained.

\section{Applicability}
\label{sec-applicability}

In the above, we have assumed that $T_R$ is fixed, limited by some
independent experimental factor.  This assumption is warranted in many
trapped-ion atomic clock experiments, where, for example, we want to limit the
heating that takes place with laser cooling radiation absent.
(During application of the Ramsey fields the cooling radiation must be removed
to avoid perturbing the clock states.)  Additionally, we may
want to lock a local oscillator to the atomic reference in a
practical time \cite{berkeland+98,bollinger+85}, thereby limiting $T_R$.

However, the use of entangled states may not be advantageous,
given other conditions.  For example, Huelga, {\em et al.} \cite{huelga+97}
assume that the ions are subject to a certain dephasing decoherence
rate (decoherence time less than the total observation time).
In this case, there is no advantage of using maximally entangled
states over unentangled states.  The reason is that
since the maximally entangled state decoheres $L$ times faster
than the states of individual atoms, when we use the
maximally entangled state, $T_R$ must be reduced by a factor of $L$
for optimum performance.  Therefore, the gain from using
the maximally entangled state is offset by the required reduced value of $T_R$.

Reference \cite{wineland+bible} discusses another case of practical interest.  
In atomic clocks, the frequency of
an imperfect ``local" oscillator, whose radiation drives the
atomic transition, is controlled by the 
atom's absorption resonance.  Depending on the spectrum of this
oscillator's frequency fluctuations (when not controlled) the use of
entangled states may or may not be beneficial.

\section{Implementations}

If we are able to create, with good fidelity, the state
$\psi_{max}$ (Eq.\ (\ref{psi-max})), how do we perform spectroscopy?
First, we note that $\psi_{max}$ is the state we want {\em after} the first
Ramsey $\pi/2$ pulse.  Therefore, if we were to follow as closely as
possible the Ramsey technique,
we would take $\psi_{max}$ and apply a $\pi/2$ pulse of
radiation at frequency $\omega_0$ to make the input state for the Ramsey
radiation.  However the first Ramsey $\pi/2$ pulse would only reverse this
step; therefore, it is advantageous to take the creation of $\psi_{max}$
as the first Ramsey $\pi/2$ pulse.
The second Ramsey pulse (after time $T_R$)
can be applied directly with
radiation at frequency $\omega_R$.
The phase of this pulse (on each ion) must be fixed relative to the
phases of the radiation used to create $\psi_{max}$.  In general,
the relation between these phases and $\phi_0$ (Eq. (\ref{psi-max})) will depend
on the relative phases of the fields at the positions of
each of the ions \cite{wineland+bible,turchette+98}.  This will lead to
a signal $S = \langle \tilde{O} \rangle$ $ \propto cos(L \Delta \omega T_R + \phi_f)$
where $\Delta \omega \equiv \omega_R - \omega_0$ and where
$\phi_f$ depends on all of these phases.

We can extract $\omega_0$ (relative to $\omega_R$) by measuring
$\langle \tilde{O} \rangle$ 
as a function of $T_R$, with $\Delta \omega$ fixed.
This can be further simplified by measuring 
the signal for two values of $T_R$, $T_{R2} \gg T_{R1}$, where 
$\langle \tilde{O} \rangle \simeq 0$.
Unfortunately, if the measured signal
has a systematic bias as a function of $T_R$,
an error in the determination of $\Delta \omega$ will
result.  This might happen, for example, if the ions heat up during
application of the Ramsey radiation and a loss of signal occurs due to a
reduced overlap between the ions and the laser used for fluorescence
detection of the states.
This problem could be overcome by measuring $\langle \tilde{O} \rangle$ 
for two values of $\omega_R$, $\omega_{R1}$ and $ \omega_{R2}$ such that 
$\omega_{R1} - \omega_0 \simeq -(\omega_{R2} - \omega_0)$ (determined by
the above method), 
and two values of $T_R$, $T_{R1} \ll T_{R2}$.  We then iterate the
following steps: 
(1) we make $\langle \tilde{O}((\omega_{R1} - \omega_0) T_{R1}) \rangle \simeq
\langle \tilde{O}((\omega_{R2} - \omega_0) T_{R1}) \rangle$ by adjusting the
phase of the final $\pi/2$ pulse to make $\phi_f \rightarrow 0$.  This
will take a negligible amount of time since $T_{R1} \ll T_{R2}$.
(2) We make $\langle \tilde{O}((\omega_{R1} - \omega_0) T_{R2}) \rangle \simeq
\langle \tilde{O}((\omega_{R2} - \omega_0) T_{R2}) \rangle$ by adjusting
$\omega_{R1}$ and/or $\omega_{R2}$ to force 
$\omega_{R1} - \omega_0 \rightarrow -(\omega_{R2} - \omega_0)$.
This gives $\omega_0$ relative to $\omega_R$ even if $\langle \tilde{O} \rangle$ has a
systematic bias as a function of $T_R$.

An alternative solution is suggested by Huelga, {\em et al}. \cite{huelga+97}.
After $T_R$, 
instead of applying a $\pi/2$ pulse of radiation at frequency $\omega_R$,
we apply the time-reversed sequence of operations which created
$\psi_{max}$.  This has the advantage of cancelling out all of the CN
phases that contribute to $\phi_0$ and maps the signal
($\propto cos(L \Delta \omega T_R)$) onto a single ion (whereupon
$S_{z}$ is measured for that ion).  This also reduces the
problem of detection efficiency to one ion rather than $L$ ions.  The
disadvantage of this technique is that for large values of $T_R$, the
motional mode used for logic will, most likely, have to be recooled.
This would require sympathetic cooling with the use of an ancillary ion
which, to avoid the decohering effects of stray light scattering on the logic ions,
might have to be another ion species \cite{wineland+bible}.

A more serious limitation to the accurate determination of $\omega_0$ is that, in
practice, $\psi_{max}$ will be realized only approximately and the state
produced by the logic operations will also be composed of states
other than the \mbox{$\su_1\su_2\cdots\su_L$} and \mbox{$\sd_1\sd_2\cdots\sd_L$}
states; these other states will have a definite phase relation to
the \mbox{$\su_1\su_2\cdots\su_L$} and \mbox{$\sd_1\sd_2\cdots\sd_L$}
states.
Consequently, in general, the signal produced with either
implemenation will be of the form

\begin{equation}
S = \sum_{p=1}^L C_p cos(p \Delta \omega T_R + \xi_p).
\end{equation}
To accurately determine $\Delta \omega$, it will be necessary to Fourier
decompose $S$.  Since this will take more measurements, the advantages of
using entangled states will be reduced.

In spite of this,
in some applications, it will be useful to 
determine changes in $\omega_0$ with respect to some external influence.
For example, we might want to detect changes in $\omega_0$ caused by changes
in an
externally applied field.  
In this case, as long
as $|C_p| \ll 1 $, for all $p < L$, we derive the benefits of entangled
states (assuming the decoherence time is longer than $T_R/L$) by measuring
changes in $S$ for a particular value of $T_R$.

\section{Experiments}
\label{sec-experiments}

As usual, our enthusiasm for implementing these schemes
far exceeds what is accomplished in
the laboratory; nevertheless, some encouraging signs are apparent from
recent experiments.  In Ref. \cite{king+98},  all motional modes for two
trapped ions have been cooled to the ground state.  The
non-center-of-mass modes are observed to be much less susceptible to
heating, suggesting the use of these modes in quantum computation
or quantum state engineering.
In Ref. \cite{turchette+98}, we describe logic
operations which enabled $\psi_{max}$ for $L=2$ to be generated with modest
fidelity ($ \simeq 0.7$).  For small $L$, it is only necessary to {\em
differentially} address individual ions to create $\psi_{max}$ and for
$L=2$, general logic can be realized even if the laser beams cannot be
focused exclusively on the individual ions 
\cite{turchette+98}.  For general logic on more than two ions, two avenues are
being pursued.  For modest numbers of ions in a
trap, the Cirac-Zoller scheme of individual addressing with the use of
focused laser beams is the most
attractive.  Current efforts are devoted to obtaining sufficiently
strong focusing to achieve individual ion addressing in a relatively
strong trap where normal mode frequencies are relatively high ($\simeq
10$ MHz) in order to maximize operation speed.
Alternatively, general logic on many ions could be accomplished
by incorporating accumulators \cite{wineland+bible},
and using differential addressing on two
ions at a time.  This idea might be realized by scaling up a version of a linear ion trap
made with lithographically deposited electrodes as we have recently
demonstrated \cite{haroche98,myatt+98}.  Concurrently, efforts are
being devoted to the investigation (and hopefully, elimination) of mode heating
\cite{wineland+bible} for different electrode surfaces and dimensions.

\section{Acknowledgments}

We gratefully acknowledge the support of the U.S. National Security
Agency, U.S. Army Research Office, and the U.S. Office of Naval
Research.  We thank J. Bollinger, R. Blatt, D. Sullivan, and M. Young
for helpful comments on the manuscript.

\end{document}